# Coevolutionary search for optimal materials in the space of all possible compounds

Zahed Allahyari[1, 2, *] and Artem R. Oganov[1, 2, 3, †]

[1] Skolkovo Institute of Science and Technology, Skolkovo Innovation Center, 3 Nobel Street, Moscow 143026, Russia

[2] Moscow Institute of Physics and Technology, 9 Institutsky Lane, Dolgoprudny 141700, Russia

[3] International Center for Materials Discovery, Northwestern Polytechnical University, Xi'an 710072, China

*zahed.allahyari@gmail.com, †A.Oganov@skoltech.ru

**Abstract**

Over the past decade, evolutionary algorithms, data mining and other methods showed great success in solving the main problem of theoretical crystallography: finding the stable structure for a given chemical composition. Here we develop a method that addresses the central problem of computational materials science: the prediction of material(s), among all possible combinations of all elements, that possess the best combination of target properties. This nonempirical method combines our new coevolutionary approach with the carefully restructured "Mendelevian" chemical space, energy filtering, and Pareto optimization to ensure that the predicted materials have optimal properties and a high chance to be synthesizable. The first calculations, presented here, illustrate the power of this approach. In particular, we find that diamond (and its polytypes, including lonsdaleite) are the hardest possible materials and that bcc-Fe has the highest zero-temperature magnetization among all possible compounds.

## 1. Introduction

Finding materials with optimal properties (e.g., the highest hardness, the lowest dielectric permittivity, etc.) or a combination of properties (e.g., the highest hardness and fracture toughness) is the central problem of materials science. Until recently, only experimental materials discovery was possible, with all limitations and expense of the trial-and-error approach, but the ongoing revolution in theoretical/computational materials science (see ref [1,2]) begins to change the situation. Using quantum-mechanical calculations, it is now routine to predict many properties when the crystal structure is known. In 2003, Curtarolo demonstrated the data mining method for materials discovery [3] by screening crystal structure databases (which can include known or hypothetical structures) via ab initio calculations. At the same time, major progress in fully nonempirical crystal structure prediction took place. Metadynamics [4] and evolutionary algorithms [5–7] have convinced the community that crystal structures are predictable. Despite the success of these and other methods, a major problem remains unsolved: the prediction of a material with optimal properties among all possible compounds. 4,950 binary systems, 161,700 ternary systems, 3,921,225 quaternary systems, and an exponentially growing number of higher-complexity systems can be created from 100 best-studied elements in the Periodic Table. In each system, a very large number of compounds and, technically, an infinite number of crystal structures can be constructed computationally, and an exhaustive screening of such a search space is impractical. Only about 72% of binary, 16% of ternary, 0.6% of quaternary, and less than 0.05% of more complex systems have ever been studied experimentally,[8] and even in those systems that have been studied, new compounds are being discovered continually.[9–11] Studying all these systems, one by one, using global optimization methods is unrealistic. Data mining is a more practical approach, but the statistics shows that the existing databases are significantly incomplete even for binary systems, and much more so for ternary and more complex systems. Besides, data mining cannot find fundamentally new crystal structures. When searching for materials optimal in more than one property, these limitations of both approaches become even greater. We present a new method implemented in our code, MendS (Mendelevian Search), and show its application to the discovery of (super)hard and magnetic materials.

## 2. Results

**Mendelevian space**

Global optimization methods are effective only when applied to property landscapes that have an overall organization, e.g., a landscape with a small number of funnels, where all or most of the good solutions (e.g., low-energy structures) are clustered. Discovering materials with optimal properties, i.e. performing a complex global optimization in the chemical and structural space, requires a rational organization of the chemical space that puts compounds with similar properties close to each other. If this space is created by ordering the elements by their atomic numbers, we observe a periodic patchy pattern (Fig. 1a), unsuitable for global optimization.

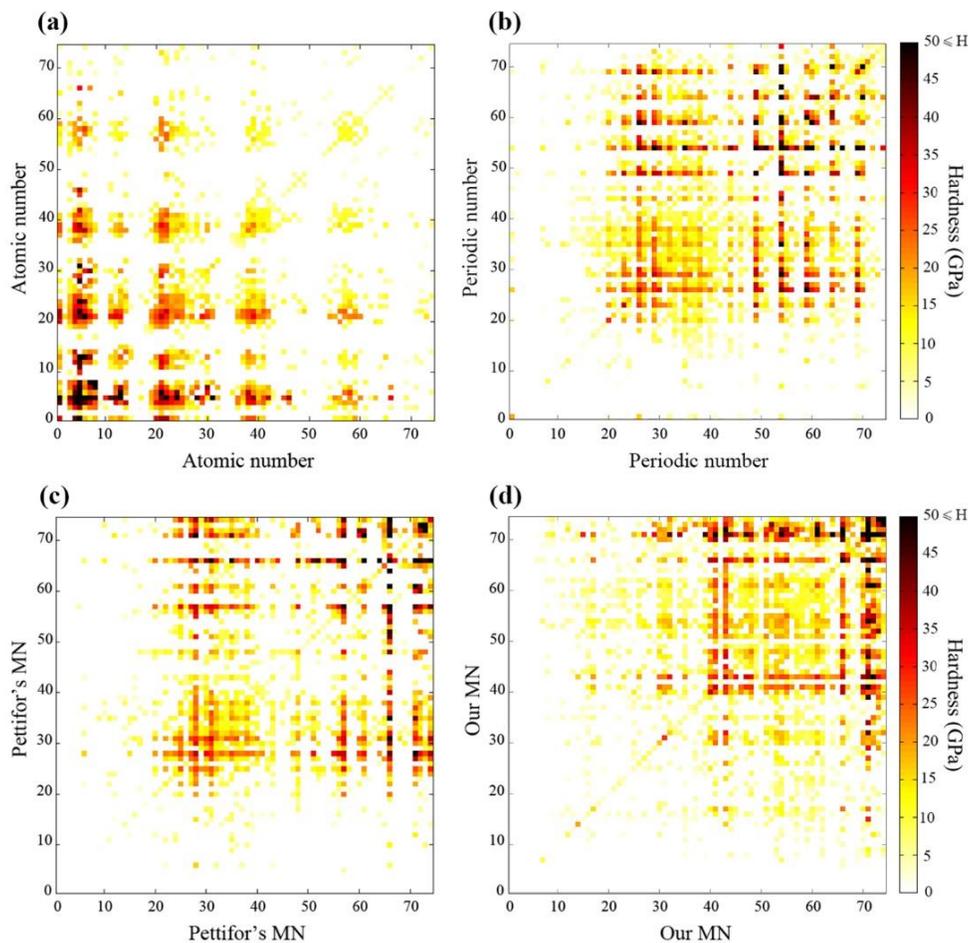

**Figure 1. Pettifor maps showing the distribution of hardness in binary systems, using different sequence of the elements.** (a) atomic numbers, (b) Villars' Periodic number, (c) Pettifor's MN, and (d) MN obtained in this work. Noble gases were excluded because of their almost complete inability to form stable compounds at normal conditions. Rare earths and elements heavier than Pu were excluded because of the problems of the DFT calculations. In total, we consider 74 elements that can be combined into 2,775 possible binary systems. Each pixel is a binary system, the color encodes the highest hardness in each system.

In 1984, Pettifor suggested a new quantity, the so-called "chemical scale," that arranges the elements in a sequence such that similar elements are placed near each other, and compounds of these elements also display similar properties.[12] This way, structure maps[13] with well-defined regions of similar crystal structures or properties can be drawn. In a thus ordered chemical space, evolutionary algorithms should be extremely effective: they can zoom in on the promising regions at the expense of unpromising ones.

What is the nature of the chemical scale or the Mendeleev number (MN), which is an integer showing the position of an element in the sequence on the chemical scale? Pettifor derived these quantities empirically, while we redefined them using a more universal nonempirical way that clarifies their physical meaning (the method for computing MN is explained in the Supplementary Information). Goldschmidt's law of crystal chemistry states that the crystal structure is determined by stoichiometry, atomic size, polarizability, and electronegativity of atoms/ions,[14,15] while polarizability and electronegativity are strongly correlated.[16] Villars et al., introduced another enumeration of the elements, emphasizing the role of valence electrons, which he called "Periodic number" (PN).[17] He also showed that atomic size and electronegativity can be derived from AN and PN.[17] In redefining the chemical scale and Mendeleev number, we used the most important chemical properties of the atom — size $R$ and electronegativity $\chi$ (Pauling electronegativity) — the combination of which can be used as a single parameter succinctly characterizing the chemistry of the element. However, we need to emphasize that the chemical scale and MN are only used in this method for visualizing the results (the choice of MN for plotting such a Pettifor map is up to the user), while in our global coevolutionary algorithm, each atom is represented by both its size $R$ and electronegativity $\chi$ to increase the accuracy. In this work, the atomic radius $R$ is defined as half the shortest interatomic distance in the relaxed (for most elements hypothetical) simple cubic structure of an element – see the Supplementary Table 1.

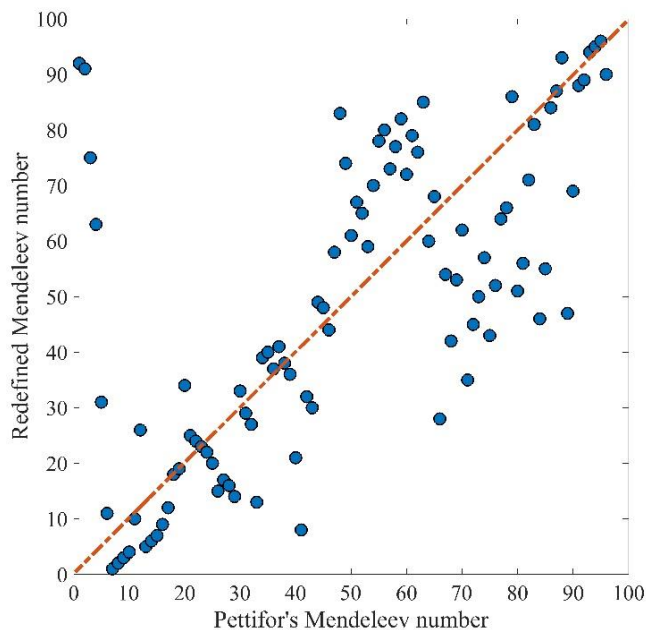

**Figure 2. Correlation between the Mendeleev numbers defined in this work and those proposed by Pettifor.** It is clear that these MNs have overall correlation, but for some elements (i.e. noble gases) there are big differences.

Fig. 2 shows the overall linear correlation between the Mendeleev numbers redefined in this work and those proposed by Pettifor. Carefully chosen Mendeleev numbers should lead to strong clustering in the chemical space, where neighboring systems have similar properties. The results of our searches for hard binary compounds using the Periodic number, the MNs suggested by Pettifor and our redefined MNs are shown on Pettifor maps (Fig. 1b, c, d). Satisfyingly, our redefined MNs result in a better-organized chemical space with a clearer separation of regions containing binary systems with similar hardness. In fact, if our MNs (which are the sequences of projected elements on their regression line in the space of crudely correlated atomic radius and electronegativity) generate a good 2D map, with clear grouping of similar chemical systems (e.g., Na-Cl, K-Cl, Ca-Cl, Na-Br systems are located nearby), then a much better grouping is expected in the space of the initial two parameters $R$ and $\chi$, and it is in this space where variation operators of our method are defined (Fig. 3a, b). Also it worth mentioning that sizes and electronegativities of the atoms change under pressure – and using standard definitions of the Mendeleev number (such as AN, PN, or Pettifor's MN) will not work well. Our recipe, however, is universal and only requires atomic sizes and electronegativities at the pressure of interest. In this paper, we illustrate our method by binary systems, although more complex, at least ternary, systems are also tractable. In a nutshell (but see Methods section for details), our method performs evolution of a population of variable-composition chemical systems (each of which is tackled by an evolutionary optimization) - i.e. is an evolution over evolutions. Individual chemical systems are allowed to evolve and improve, then are compared and ranked, and the fittest ones get a chance to produce new chemical systems (which will partially inherit structural and chemical information from their parents). Evolving the population of such chemical systems, one efficiently finds the globally optimal solution and numerous high-quality suboptimal solutions as well.

**Search for hard and superhard binary systems**

Pareto optimization[18] of hardness and stability was performed over all possible structures (with up to 12 atoms in the primitive cell) and compositions limited to the binary compounds of 74 elements (i.e. all elements excluding the noble gases, rare earth elements, and elements heavier than Pu). In this work, 600 systems have been computed in 20 MendS generations from a total of 2,775 unary and binary systems that can be made of 74 elements, i.e. only about one fifth of all possible systems were sampled.

Fig. 4 shows the efficiency of this method in finding optimal materials. In this fast calculation, numerous stable and metastable hard and superhard materials were detected in a single run. Carbon (diamond and other allotropes) and boron, known to be the only superhard elements, were both found. In addition, both new and numerous known hard and superhard binary systems, as well as potentially hard systems, were found in the same calculation, among them $B_xC_y$,[19] $C_xN_y$,[20,21] $B_xN_y$,[22,23] $B_xO_y$,[19,24,25] $Re_xB_y$,[26,27] $W_xB_y$,[28] $Si_xC_y$,[29–32] $W_xC_y$,[30–32] $Al_xO_y$,[30–32] $Ti_xC_y$,[32] $Si_xN_y$,[32] $Ti_xN_y$,[32] $Be_xO_y$,[32] $Ru_xO_y$,[33,34] $Os_xO_y$,[35] $Rh_xB_y$,[36] $Ir_xB_y$,[36] $Os_xB_y$,[37–39] and $Ru_xB_y$.[37–39] We reported some of the results of our search in a separate paper on the Cr-B, Cr-C, and Cr-N systems,[40] and our study of the W-B system[41] was inspired by the present finding of promising properties in the Mo-B system (also published in [42]). The list of all systems studied during the calculation is available in Supplementary Information.

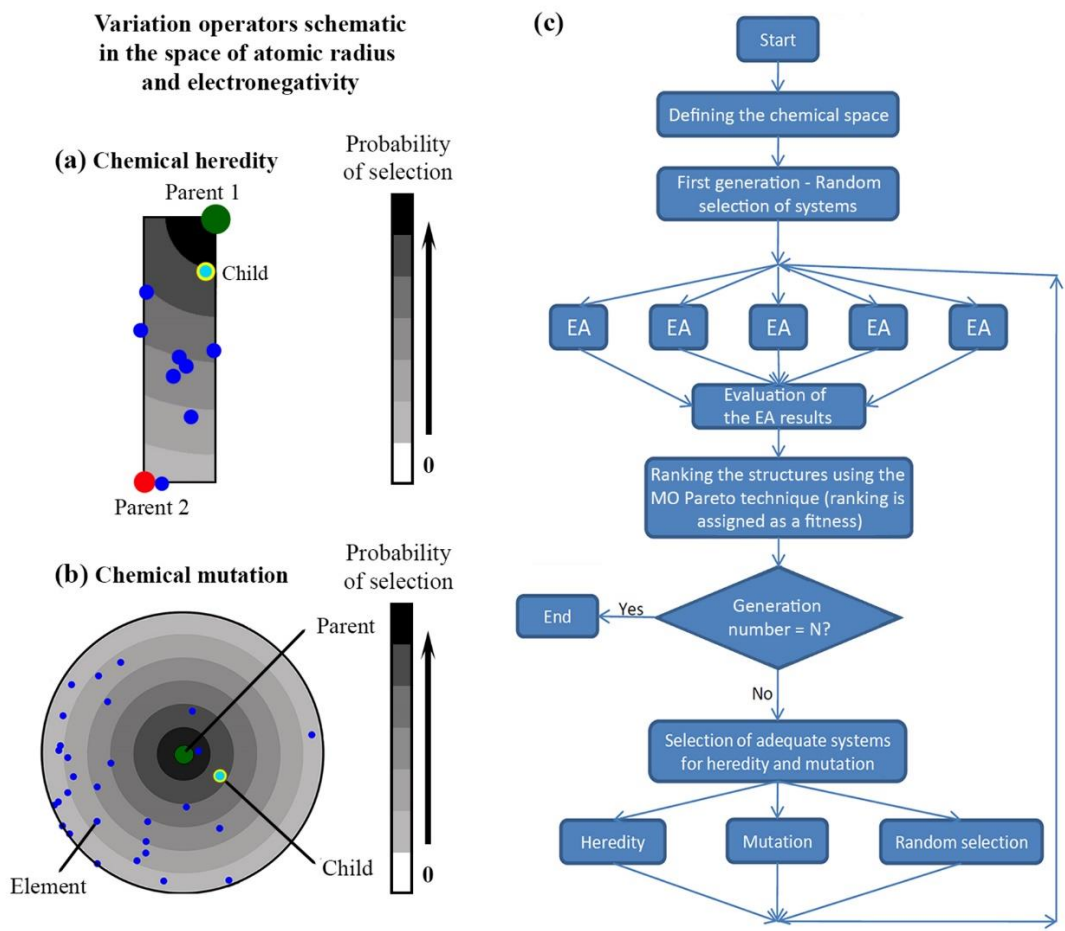

**Figure 3. MendS algorithm**, (a) Scheme showing how the chemical heredity and (b) chemical mutation create new compositions. The probability, displayed in shades of gray, is given to each possible child according to its distance from the fitter parent (dark green point). (c) Flowchart of the coevolutionary algorithm used in MendS (EA — evolutionary algorithm, MO — multi-objective).

Because of the huge compositional space (2,775 systems, each with $10^2$ possible compositions, each of which having a very large number of possible structures), it was necessary to shorten the time of calculations by reducing the number of generations and/or population size. Therefore, the structures and compositions found may be approximate and may need to be refined for the most interesting systems by a precise evolutionary calculation for each system. The results are shown in Table 1. Of these, some transition metal borides are predicted to be hard, some already reported as hard materials (e.g., $Mo_xB_y$,[43,44] $Mn_xB_y$[45]) or discussed as potentially hard (e.g., $Tc_xB_y$,[46] $Fe_xB_y$,[47] and $V_xB_y$[48]). Interestingly, a number of previously unknown hard structures more stable than those reported so far were predicted in these systems. Our calculations also revealed completely new hard systems, $S_xB_y$ and $B_xP_y$, and, quite unexpectedly, the $Mn_xH_y$ system was discovered to contain very hard phases (Table 1).

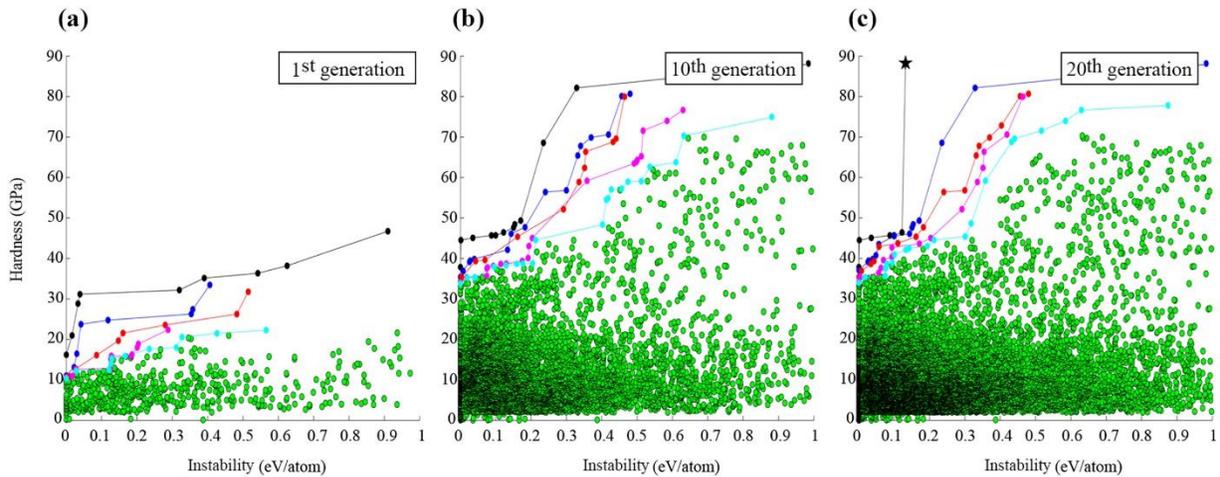

**Figure 4. Results of the simultaneous optimization of hardness and stability in the space of all unary and binary compounds:** (a) 1st MendS generation, (b) 10th MendS generation, (c) 20th MendS generation. The first five Pareto fronts are shown, green points representing all sampled structures. The instability of each compound is defined using Maxwell's convex hull construction. Diamond, the hardest material, is indicated by a star.

**Table 1. The predicted Vickers hardness ($H_v$), fracture toughness ($K_{IC}$) and enthalpy above the convex hull of selected materials found using MendS.** Theoretical values from previous works are shown in parentheses, experimental values are in brackets. The values of hardness for superhard materials (harder than 40 GPa) are highlighted in bold. The hardness was computed using the Chen-Niu model,[49] the fracture toughness — using the Niu-Niu-Oganov model.[50] Ref: a [29], b [30], c [43], d [44], e [51], f [52], g [53], h [54], i [55], j [56], k [57], m [58], n [59], p [60], q [61].

| | Compounds | $H_v$ (GPa) | $K_{IC}$ (MPa·m$^{1/2}$) | Instability (eV/atom) | Space group | | Compounds | $H_v$ (GPa) | $K_{IC}$ (MPa·m$^{1/2}$) | Instability (eV/atom) | Space group |
|---|---|---|---|---|---|---|---|---|---|---|---|
| Carbon | diamond | **92.7** (**93.6**)[b] [**96**][b] | 6.33 | 0.13 | $Fd\bar{3}m$ | Boron | α-boron | 38.9 (39)[h] [27-34][i] | 2.87 | 0 | $R\bar{3}m$ |
| | lonsdaleite | **93.6** | 6.36 | 0.139 | $P6_3/mmc$ | | B | **44.8** | 3.29 | 0.136 | $Cmc2_1$ |
| B-S | $B_4S_3$ | 30.5 | 1.83 | 0.102 | $Cmcm$ | B-N | BN | **63.4** (**64.5**)[b] [**66**][b] | 5.1 | 0.075 | $F\bar{4}3m$ |
| Mo-B | $MoB_2$ | 28.5 (33.1)[d] [24.2][e] | 3.76 | 0 | $R\bar{3}m$ | Tc-B | TcB | 31 (30.3)[j] | 3.83 | 0.013 | $P\bar{3}m1$ |
| | $MoB_3$ | 35.3 | 3.74 | 0.035 | $P\bar{3}m1$ | | $TcB_3$ | 27.2 (29)[k] | 3.6 | 0 | $P\bar{6}m2$ |
| | $MoB_3$ | 32.2 | 3.63 | 0.077 | $A2/m$ | | $TcB_3$ | 33.1 | 3.79 | 0.003 | $P\bar{3}m1$ |
| | $MoB_3$ | 35.3 (37.3)[d] | 3.63 | 0.017 | $P6_3/mmc$ | | $TcB_4$ | 31.8 | 3.56 | 0.069 | $P2_1/m$ |
| | $MoB_3$ | 33.1 (31.8)[c] | 3.57 | 0.011 | $R\bar{3}m$ | | $TcB_4$ | 30.2 | 3.54 | 0.069 | $R\bar{3}m$ |
| | $MoB_4$ | 35.4 | 3.57 | 0.099 | $Pmmn$ | | $TcB_4$ | 30 (32)[k] | 3.57 | 0.027 | $P6_3/mmc$ |
| | $MoB_5$ | 35.7 | 3.62 | 0.054 | $P\bar{6}m2$ | | $TcB_7$ | 35.9 | 3.35 | 0.084 | $R3m$ |
| | $MoB_8$ | 36.6 | 3.24 | 0.118 | $R3m$ | | $TcB_8$ | 33.9 | 3.3 | 0.113 | $R3m$ |
| | $Mo_2B_3$ | 32.2 | 3.95 | 0.029 | $Imm2$ | | $Tc_3B_5$ | 30.6 | 3.87 | 0 | $P\bar{6}m2$ |
| | $Mo_2B_3$ | 30.4 | 3.87 | 0.043 | $Cmcm$ | | | | | | |
| Si-C | SiC | 33.3 (33.1)[a] [34][b] | 2.94 | 0 | $F\bar{4}3m$ | B-P | BP | 37.2 (31.2)[b] [33][b] | 2.46 | 0 | $F\bar{4}3m$ |
| | SiC | 33.1 | 2.94 | 0.001 | $R3m$ | | $B_6P$ | **41.1** | 2.87 | 0 | $R\bar{3}m$ |
| V-B | VB | 39.1 (38.3)[m] | 3.66 | 0 | $Cmcm$ | Mn-H | MnH | 29.5 | 3.2 | 0 | $P6_3/mmc$ |
| | $VB_2$ | 37.3 (39.5)[m] [27.2][n] | 3.75 | 0 | $P6/mmm$ | | MnH | 27.9 | 3.14 | 0.013 | $R\bar{3}m$ |
| | $VB_5$ | **40** | 3.36 | 0.158 | $P\bar{6}m2$ | | MnH | 26.3 | 3.07 | 0.044 | $Fm\bar{3}m$ |
| | $VB_7$ | 39.7 | 3.19 | 0.143 | $P3m1$ | | $Mn_3H_2$ | 26.8 | 3.22 | 0.017 | $R32$ |
| | $VB_{12}$ | **44.5** | 3.34 | 0.125 | $I4/mmm$ | | $Mn_3H_2$ | 27 | 3.26 | 0.019 | $P6_3/mcm$ |
| | $V_3B_4$ | 37.8 | 3.74 | 0 | $P\bar{4}m2$ | | $Mn_4H_3$ | 27.6 | 3.23 | 0.002 | $P2/m$ |
| | $V_3B_4$ | 35.9 (38.2)[m] | 3.7 | 0.006 | $Immm$ | | $Mn_6H_5$ | 27.3 | 3.17 | 0.011 | $A2/m$ |
| Mn-B | $MnB_3$ | 32.2 | 3.5 | 0.029 | $P\bar{6}m2$ | Fe-B | $FeB_3$ | 30.2 | 3.32 | 0 | $P2_1/m$ |
| | $MnB_4$† | **40.7** | 3.65 | 0.009 | $Pnnm$ | | $FeB_4$ | 35.7 | 3.06 | 0.021 | $Immm$ |
| | $MnB_4$ | 38.2 | 3.56 | 0.1 | $R\bar{3}m$ | | $FeB_4$‡ | 32 | 3.31 | 0.039 | $R\bar{3}m$ |
| | $MnB_4$ | 38.1 (**40.5**)[f] [37.4][g] | 3.76 | 0 | $P2_1/c$ | | $FeB_4$ | **42.7** | 3.31 | 0.063 | $A2/m$ |
| | $MnB_5$ | 32.7 | 3.38 | 0.097 | $P\bar{6}m2$ | | $FeB_4$ | 28.6 (28.4)[p] [**62**][q] | 3.32 | 0.002 | $Pnnm$ |
| | $MnB_{13}$ | **40.4** | 2.9 | 0.181 | $Pm$ | | $Fe_2B_{11}$ | 33.8 | 3.37 | 0.081 | $Pm$ |

†‡ For these phases we found that ferromagnetic solutions are more stable than non-magnetic. Elastic constant were computed assuming these are ferromagnetic structures, the energy difference between the ferromagnetic and non-magnetic solutions for † and ‡ is 0.037 (eV/transition-metal) and 0.092 (eV/transition-metal) and magnetization is equal to 0.016 and 0.034 $\mu_B·Å^{-3}$, respectively.

For the $Mo_xB_y$ system, several simultaneously hard and low-energy structures were detected in our calculations. Of these, only the stable $R\bar{3}m$ structure of $MoB_2$ was studied before, and the reported hardness for this structure (experimentally obtained 24.2 GPa[51] and theoretically found 33.1 GPa[44]) is in close agreement with the value calculated in this work (28.5 GPa). $MoB_3$ and $MoB_4$ were studied widely before,[43,44] and a few low-energy and in some cases hard structures were reported for these systems (i.e. $R\bar{3}m$-$MoB_3$, 31.8 GPa[43]; $P6_3/mmc$-$MoB_3$, 37.3 GPa[44]; and much softer $P6_3/mmc$-$MoB_4$, 8.2 GPa[44]). In this work, new low-energy structures with high hardness were discovered in these systems (Table 1).

For the $Mn_xB_y$ system, we propose several new compounds which are simultaneously hard and have low energy (Table 1). In the previous study,[52] $P2_1/c$-$MnB_4$ was discussed to be stable and have a very high hardness (computed to be 40.1 GPa,[52] experimentally obtained 34.6-37.4 GPa[53]), while $C2/m$-$MnB_4$ was claimed to be the second lowest-energy structure with the energy difference of 18 meV/atom. Our study confirms the stability of $P2_1/c$-$MnB_4$. However, we discovered another $MnB_4$ structure, with the *Pnnm* space group, whose energy lies between the energies of two aforementioned structures of $MnB_4$ (Table 1). In this work, we found that the ferromagnetic phase of *Pnnm*-$MnB_4$ is more stable than the nonmagnetic one, and the hardness of 40.7 GPa was computed for this magnetic structure.

Because of the radioactivity of technetium, the $Tc_xB_y$ system has not been studied experimentally, while computational studies of this system started recently.[46,54,55,60] In 2015, $P\bar{3}m1$-TcB was predicted to be energetically more favorable than the previously discussed *Cmcm* and WC-type structures.[56] The reported hardness for this structure, 30.3 GPa,[56] is very close to the value predicted in this study (31 GPa). Because of the prediction of other stable compounds (e.g., $Tc_3B_5$) in our work, this structure became metastable (by 13 meV/atom above the convex hull). In this work, $P\bar{6}m2$-TcB$_3$ with the computed hardness of 27.2 GPa was predicted as a stable structure at zero pressure. Other works,[55,60] conducted in parallel to ours, also detected this structure and claimed that it is synthesizable at pressures above 4 GPa.[57] Another low-energy (3 meV above the convex hull) hard structure (33.1 GPa) with the $P\bar{3}m1$ space group for TcB$_3$ was also predicted in our study. $P\bar{6}m2$-Tc$_3$B$_5$, a compound having a hardness of 30.6 GPa and stable at zero pressure, is predicted in our work for the first time. Several other simultaneously hard (in the range of 30 to 36 GPa) and low-energy metastable phases of Tc$_x$B$_y$ predicted in this work are shown in Table 1.

In recent years, many efforts were focused on searching for low-energy phases of V$_x$B$_y$ and studying their electrical and mechanical properties. As a result, several low-energy hard and superhard phases were predicted.[48,56] Nevertheless, the experimental data exist only for the well-known hexagonal VB$_2$ (AlB$_2$-type) with the $P6/mmm$ space group.[59] In addition to some previously studied structures[58] (e.g., $Cmcm$-VB, $Immm$-V$_3$B$_4$, and $P6/mmm$-VB$_2$), which were also found in our calculations, a few boron-rich phases possessing simultaneously low energy and very high hardness were discovered (Table 1). The calculated hardness for these boron-rich phases is very close to or above 40 GPa (VB$_7$: 39.7 GPa, VB$_5$: 40 GPa, and VB$_{12}$: 44.5 GPa). A new extremely hard $P\bar{4}m2$-V$_3$B$_4$ phase is predicted here, with the energy 6 meV lower than the previously proposed $Immm$ structure.

Most of the studies of the Fe$_x$B$_y$ system were dedicated to the FeB$_2$ and FeB$_4$ phases.[47,61,62] Several works studying different Fe$_x$B$_y$ compounds[63,64] reported Fe$_2$B, FeB, and FeB$_2$ as stable phases. In this work, we detected another stable phase, FeB$_3$, with the $P2_1/m$ space group and the hardness of 30.7 GPa. To the best of our knowledge, FeB$_3$ was never reported, neither theoretically nor experimentally. The orthorhombic $Pnnm$-FeB$_4$, with the energy of 2 meV above the convex hull (Table 1), was synthesized at pressures above 8 GPa, and its hardness was reported to be 62(5) GPa,[61] which encouraged many computational studies of this structure. However, none of them confirmed such a high value of hardness, while the Vickers hardness reported in several independent works varies in the range of 24 to 29 GPa.[47,62,64,65] We calculated its hardness to be 28.6 GPa.

In the $B_xP_y$ system, the cubic boron phosphide BP with the zincblende structure is a well-known compound with the hardness reported to be roughly the same as that of SiC.[66] In our calculations, the hardness of SiC and BP was found to be 33 GPa and 37 GPa, respectively. Moreover, $B_6P$ was discovered as another stable compound in this system and predicted to be superhard, with the computed hardness exceeding 41 GPa. In the SiC system, in addition to the knowndiamond-type β-SiC, another similar structure (actually, a polytype of β-SiC) with the *R3m* space group and nearly the same hardness was found. The energy of this structure is just 1 meV/atom higher than that of β-SiC.

The $Mn_xH_y$ system is unexpected in the list of hard systems, but several very hard phases were indeed found in it (Table 1). All of these phases are nonmagnetic, highly symmetric, and energetically favorable (lying either on the convex hull or close to it), with the hardness of up to 30 GPa. In this system, two thermodynamically stable compounds ($Mn_2H$ and MnH) were predicted, with the space groups $P\bar{3}m1$ and $P6_3/mmc$, and computed hardness of 21.5 GPa and 29.5 GPa, respectively (in Table 1, only structures with the hardness above 26 GPa are shown for this system).

Generally, $B_xS_y$ system is not hard, but metastable boron sulfides turn out to be potentially hard. We found a low-energy metastable phase of this system, *Cmcm*-$B_4S_3$, with the hardness unexpectedly exceeding 30 GPa. This can stimulate future studies of this system.

For a better insight, some of the prominent structures seen in our simulations are shown in Fig. 5a. More details on all phases presented in Table 1 are given in Supplementary Information.

In our calculations, some boron hydrides were predicted to be superhard, but they had high energy and were not included in Table 1. However, it may be possible to stabilize these hard phases under pressure, or by chemical modification.

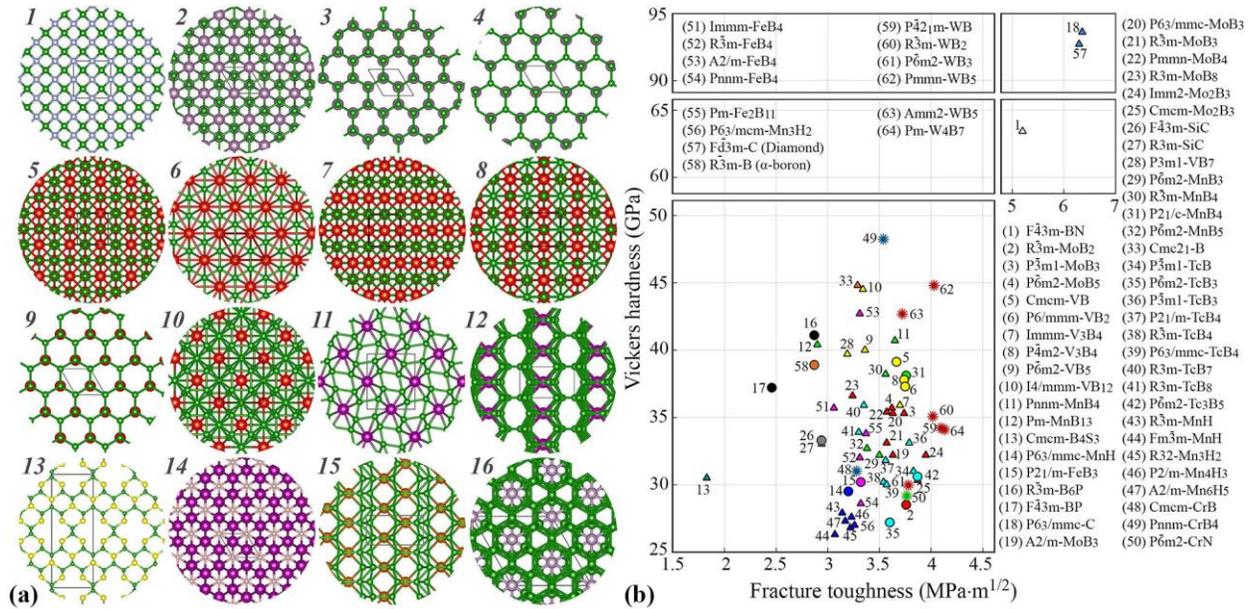

**Figure 5. Results of our Mendelevian search for hard and superhard materials.** (a) (1) $F\bar{4}3m$-BN, (2) $R\bar{3}m$-MoB$_2$, (3) $P\bar{3}m1$-MoB$_3$, (4) $P\bar{6}m2$-MoB$_5$, (5) $Cmcm$-VB, (6) $P6/mmm$-VB$_2$, (7) $Immm$-V$_3$B$_4$, (8) $P\bar{4}m2$-V$_3$B$_4$, (9) $P\bar{6}m2$-VB$_5$, (10) $I4/mmm$-VB$_{12}$, (11) $Pnnm$-MnB$_4$, (12) $Pm$-MnB$_{13}$, (13) $Cmcm$-B$_4$S$_3$, (14) $P6_3/mmc$-MnH, (15) $P2_1/m$-FeB$_3$, (16) $R\bar{3}m$-B$_6$P. (b) "Ashby plot" of the Vickers hardness vs. fracture toughness. Stable hard compounds from the previous works [40,50] are shown as suns; stable and metastable compounds found in this work are represented by circles and triangles, respectively.

Fig. 5b shows the studied materials in the space of "hardness — fracture toughness." Diamond, lonsdaleite and cubic BN possess the best properties, but are metastable at normal conditions. Among the stable phases, borides of transition metals (especially from groups VB, VIB, VIIB) stand out: we note $VB_2$, $V_3B_4$, $MoB_2$, $CrB_4$, $WB_5$, and $MnB_4$ in particular. These and related materials (see [1]) present a high technological interest.

The fact that all known binary superhard systems were found in a short coevolutionary run demonstrates the power of the method, which is ready to be applied to the other types of materials.

**Search for magnetic binary systems**

In addition to the Mendelevian search for stable/metastable hard and superhard materials, we performed another Mendelevian search for materials with maximum magnetization and stability to examine the power and efficiency of the method in fast and accurate determination of materials with target properties. We performed this calculation over all possible structures (with up to 12 atoms in the primitive cell) and compositions limited to the binary compounds of 74 elements (i.e. all elements excluding the noble gases, rare earth elements, and elements heavier than Pu). In this calculation, well-known ferromagnets iron, cobalt, nickel, and several magnetic materials made from the combination of these elements with other elements were detected before the 6th generation. Here, for each structure we performed spin-polarized calculations using the GGA-PBE functional[67] as implemented in the VASP code.[68,69] More details on structure relaxation and input parameters can be found in Supplementary Information. The chemical landscape of magnetization and evolution of its sampling in the Mendelevian search for magnetic materials are shown in Fig. 6d-f; this was formed after calculating 450 binary systems over 15 generations. In this plot, materials with high magnetization are clearly clustered together. Fig. 6d,f shows how the (co)evolutionary optimization discovered all the promising regions at the expense of the unpromising ones. This calculation has found that among all substances, bcc-Fe has the highest magnetization at zero Kelvin.

## 3. Discussion

We have developed a method for predicting materials having one or more optimal target properties. The method, called Mendelevian search (MendS), based on the suitably defined chemical space, powerful coevolutionary algorithm, and multi-objective Pareto optimization technique, was applied to searching for low-energy hard and superhard materials. Note that due to the property of evolutionary and coevolutionary algorithms to enhance sampling of the most promising regions of the search space (where the optimal, as well as all or most of the high-quality sub-optimal solutions are clustered together), each MendS search discovers a large number of materials with excellent properties at a low computational cost. Well-known superhard systems — diamond, boron allotropes, and the B-N system — were found in a single calculation together with other notable hard systems (Si-C, B-C, Cr-N, W-C, metal borides, etc.). The Mn-H system was discovered to be unexpectedly hard, and several new hard and superhard phases were revealed in the previously studied systems (V-B, Tc-B, Mn-B, etc.). The method successfully found almost all known hard systems in a single run, and a comprehensive chemical map of hard materials was produced. A similar chemical map was plotted for magnetic materials; well-known magnetic systems such as Ni, Co, Fe were found within just a few generations. These examples show the power and efficiency of our method, which can be used to search for optimal materials with any combination of properties at arbitrary conditions. As the first step in prediction of novel materials possessing desired properties, the method to a large extent solves, in a fully nonempirical way, the central problem of computational materials science.

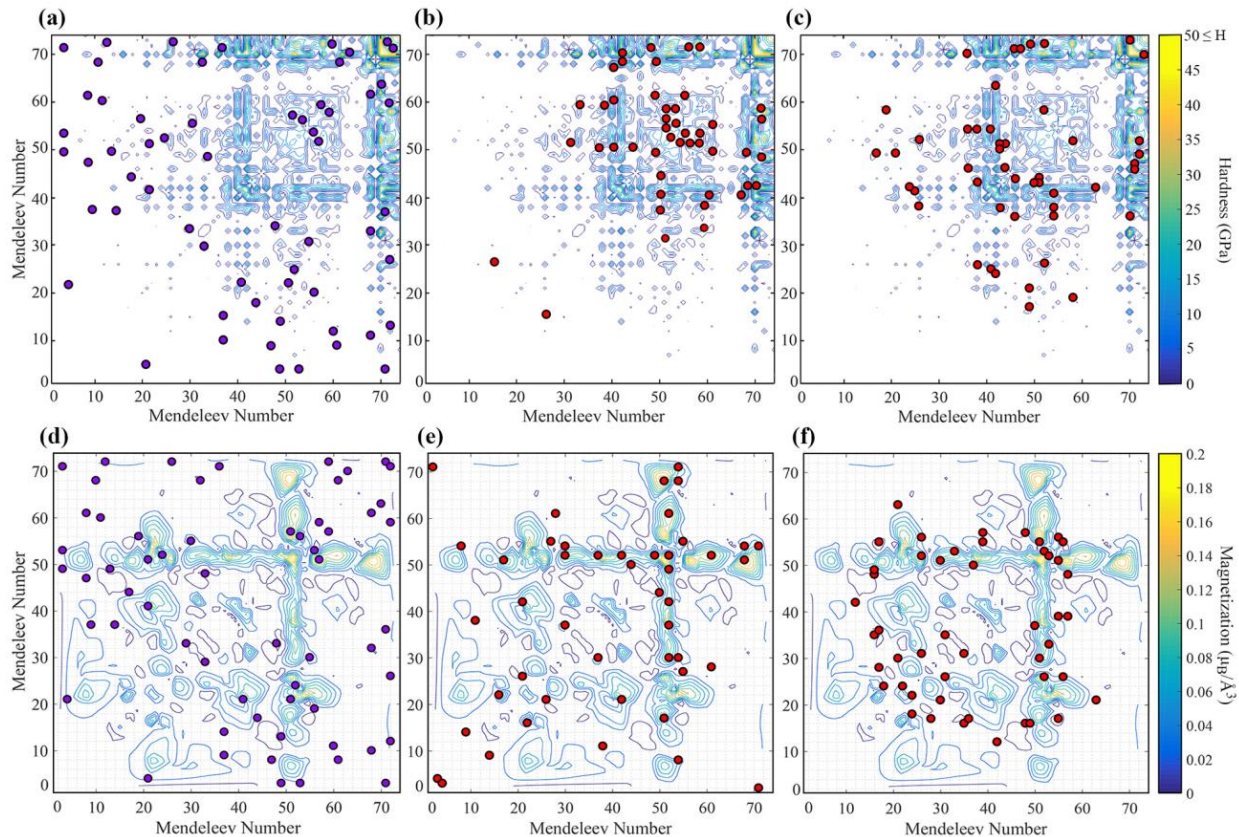

**Figure 6. Sampling of the chemical space.** Systems produced (a, d) randomly in the 1st generation, and using all variation operators in the (b, e) 5th and (c, f) 10th generations in searching for hard (a-c) and magnetic (d-f) materials. Randomly generated systems are shown as violet circles.

## 4. Methods

The whole process can be described as a joint evolution (or coevolution) of evolutionary runs, each of which deals with an individual variable-composition system. Having defined the chemical space, we initialize the calculation by randomly selecting a small number of systems from the entire chemical space for the first MendS generation. These systems are then optimized by the evolutionary algorithm USPEX [5–7] in its variable-composition mode,[70] searching for compounds and structures with optimal properties (e.g., here we simultaneously maximized hardness and stability), after which MendS jointly analyses the results from all these systems. Removing identical structures using the fingerprint method,[71] jointly evaluating all systems, refining and preparing the dataset, and discarding the structures that are unstable by more than 1.0 eV/atom, MendS ranks all systems of the current generation and selects the fittest (in present calculations, fittest 60% were selected) variable-composition systems as potential parents for new systems. Applying chemical variation operators, such as mutation and heredity, to these parent systems yields offspring systems for the next coevolutionary generation. Additionally, some systems are generated randomly to preserve the chemical diversity of the population. This process is continued until the number of coevolutionary generations reaches the maximum predefined by the user (Fig. 3c). The underlying ab initio structure relaxations and energy calculations were performed using density functional theory with the projector augmented wave method (PAW) as implemented in the VASP code.[68,69] Further details on the input parameters of MendS, USPEX, and VASP are given in Supplementary Information.

**Defining fitness: Multi-objective (Pareto) optimization**

Many scientific and engineering problems involve optimization of multiple conflicting objectives, for example, predicting novel materials that improve upon all critical properties of the known ones. The multi-objective evolutionary algorithm (MOEA) enables searching simultaneously for materials with multiple optimal properties, such as the enthalpy, hardness, density, dielectric permittivity, magnetization, etc. Here we performed searches optimizing simultaneously (1) stability, measured as the distance above the convex hull (chances of a compound to be synthesizable are higher if the compound is stable or low-energy metastable, i.e. is on the convex hull or close to it), and (2) hardness, computed using the Lyakhov-Oganov model.[72]

Hardness is a complicated property of materials which cannot be evaluated directly and rigorously from the crystal structure because it usually includes many nonlinear and mesoscopic effects. However, there are number of empirical models making it possible to estimate hardness from atomic-scale properties. The Chen-Niu empirical model [49] is based on the relation between the elastic moduli and hardness. Although this model is reliable, calculating the elastic constants of materials on a large scale is computationally expensive. A similar model based on the elastic moduli was recently proposed by Mazhnik and Oganov [73] and unlike Chen's model, does not overestimate the hardness value of materials with low or negative Poisson's ratio while for other materials gives similar results. The Lyakhov-Oganov model,[72] which computes the hardness from bond hardnesses, is more convenient for high-throughput searches: it is numerically stable, usually reliable, and can be used in calculations without significant cost, taking the crystal structure and chemical composition as input. For better understanding of the reliability of the mentioned models, a comparison of the computed values and experimental results for hardness of various materials is presented in Ref.[74]

The result of the multi-objective optimization is, in general, not a single material, but a set of materials with a trade-off between their properties, and these optimal materials form the so-called first Pareto front. Similarly, 2nd, 3rd, ... nth Pareto fronts can be defined (Fig. 4). In our method, the Pareto rank [18] is used as a fitness.

**Variation operators in the chemical space** are of central importance for an efficient sampling of the chemical space using the previously sampled compositions and structures. These operators ensure that different populations not only compete, but also exchange information, i.e. learn from each other. An efficient algorithm could be constructed where the chemical space is defined by just one number for each element — the Mendeleev number (or chemical scale); we use this for plotting the Pettifor maps, but within the algorithm itself, we resort to an even better option where each element is described by two numbers — electronegativity $\chi$ and atomic radius $R$, rescaled to be between 0 and 1 — and it is this space where the variation operators act. There are three variation operators defined in the chemical space: chemical heredity, reactive heredity, and chemical mutation.

Chemical heredity replaces elements in parent systems with new elements such that their electronegativities and atomic radii lie in-between those of their parents (Fig. 3a). In doing so, we explore the regions of the chemical space between the parents:

$$AB + CD \rightarrow XY \tag{1}$$

where *A, B, C, D, X, and Y* are different elements, *X* is between *A* and *C* or *A* and *D* which is chosen randomly, and *Y* is between the other two elements (*B* and *C* or *B* and *D*).

Reactive heredity creates offspring by taking combinations of the elements from parents. For example, if the parents are *A-B* and *C-D*, their child is one of the *A-C, A-D, B-C,* and *B-D* systems.

Chemical mutation randomly chooses one of the elements of a parent and substitutes it with an element in its vicinity in the space of $\chi$ and $R$ (Fig. 3b).

In both chemical mutation and chemical heredity, all elements are assigned the probability:

$$P_i = \frac{e^{-\alpha x_i}}{\sum e^{-\alpha x_i}}, \qquad i = 1, 2, \ldots \tag{2}$$

to be selected, where *x* is the distance of element *i* from the parent element (in the case of chemical heredity, this formula is used to give a higher weight to the fitter parent, shown by a dark green point in Fig. 3a), and α is a constant (α = 1.5 is used here). The result of applying these chemical variation operators is shown in Fig. 6: the promising regions of the chemical space are sampled more thoroughly at the expense of the unpromising regions. When a new system is produced from parent system(s), it inherits from them a set of optimal crystal structures which are added to the first generation, greatly enhancing the learning power of the method.

After finishing the coevolutionary simulation, we took the most promising systems identified in it and performed longer evolutionary runs for each of them, calculating the final hardness using the Chen-Niu model,[49] and fracture toughness — using the Niu-Niu-Oganov model.[50]

## Acknowledgments

We thank the Russian Science Foundation (grant 19-72-30043) for financial support. All the calculations were done using the supercomputer Rurik at the Moscow Institute of Physics and Technology.

## Author contributions

A.R.O. created the ideas behind the MendS method and the new Mendeleev numbers. Z.A. implemented the MendS algorithm and Pareto optimization, and performed the calculations. Z.A. and A.R.O. wrote the paper.

# Supplementary Information

**Multi-objective optimization and system selection details**

Pareto optimization technique is implemented and used in our calculation to simultaneously perform optimization of more than one property. This would help the calculation to focus on the optimal area of the search space. In general, from the total set of structures $\{S_n\}$ (n=1, .., N), and optimizing (let's say minimizing) multiple properties $f_m$ (m=1, .., M), we say structure $S_i$ dominates structure $S_j$ (i ≠ j) in terms of properties $f_k$, if $f_k(S_i) \leq f_k(S_j)$ for all properties, and $f_k(S_i) < f_k(S_j)$ in at least one of the properties. The set of structures not dominated by any other structures is, therefore, the set of Pareto optimal structures $\{S_t\}$ (t =1, .., T and T ≤ N), and creates the first Pareto front. The set of these $\{S_t\}$ structures are then removed from the set of N individuals and the cycle is repeated to find successive Pareto fronts 2, 3,… until all N structures are classified. The set of structures belonging to the first Pareto front are assigned the highest rank, those in the second Pareto front the second highest rank, and so forth.

Selection of good parents is very important for the success of the algorithm. For selection of good parents, the roulette wheel selection method[1] is used and to each system the weight function $w_r > 0$ (r = 1, 2,.., N), derived from its fitness rank, is assigned. The probability of selection of each system with fitness rank $r$ is defined as follow:

$$P_r = \frac{w_r}{n_r \times \sum w_r} \quad , \quad r = 1, 2, \dots, N \quad (1)$$

$$w_r = (N + 1 - r)^2 \quad (2)$$

where $N$ is the last fitness ranking number, which is equal to the number of all systems in simple ranking and equal to the number of Pareto fronts in multi-objective Pareto ranking. $n_r$ is equal to the number of systems having the same fitness ranking $r$.

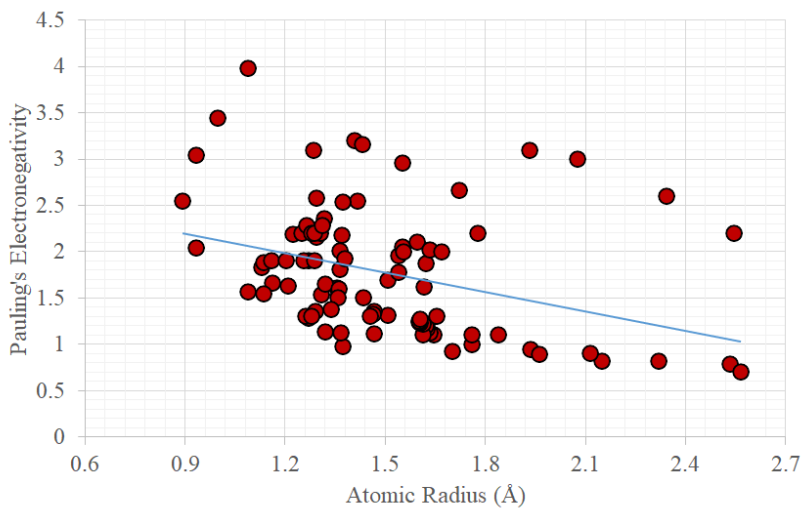

Fig 1. The electronegativities and atomic radii of the elements. The regression line is shown in blue.

## Redefining the Mendeleev Number

Unlike Pettifor, who derived his MN empirically, we offer a nonempirical (and, therefore, more universal) definition. The most important chemical properties of an atom are (1) size $R$, (2) electronegativity $\chi$, (3) polarizability $\alpha$ and (4) valence $v$. For simplicity, we excluded the valence, which is not constant for many elements. The polarizability was also disregarded in favor of the electronegativity because they are strongly correlated.[2] Thus, we only consider the electronegativity and atomic radius to define the MNs. In this work, the atomic radius $R$ is defined as half the shortest interatomic distance in the relaxed (for most elements hypothetical) simple cubic structure of an element – see Table 1.

Table 1. Atomic radii of the elements that calculated and used in this work for redefining MN.

| Element | Atomic radius $R_a$ (Å) | Element | Atomic radius $R_a$ (Å) | Element | Atomic radius $R_a$ (Å) | Element | Atomic radius $R_a$ (Å) |
|---|---|---|---|---|---|---|---|
| H | 0.727 | Mn | 1.136 | In | 1.541 | Ta | 1.358 |
| He | 1.286 | Fe | 1.131 | Sn | 1.541 | W | 1.316 |
| Li | 1.374 | Co | 1.137 | Sb | 1.553 | Re | 1.287 |
| Be | 1.090 | Ni | 1.160 | Te | 1.596 | Os | 1.278 |
| B | 0.943 | Cu | 1.203 | I | 1.721 | Ir | 1.288 |
| C | 0.891 | Zn | 1.320 | Xe | 2.344 | Pt | 1.311 |
| N | 0.932 | Ga | 1.365 | Cs | 2.535 | Au | 1.374 |
| O | 0.997 | Ge | 1.365 | Ba | 1.962 | Hg | 1.556 |
| F | 1.089 | As | 1.369 | La | 1.647 | Tl | 1.617 |
| Ne | 1.409 | Se | 1.418 | Ce | 1.467 | Pb | 1.622 |
| Na | 1.701 | Br | 1.551 | Pr | 1.367 | Bi | 1.635 |
| Mg | 1.508 | Kr | 2.077 | Nd | 1.320 | Po | 1.670 |
| Al | 1.355 | Rb | 2.319 | Pm | 1.635 | At | 1.777 |
| Si | 1.269 | Sr | 1.935 | Sm | 1.626 | Rn | 2.544 |
| P | 1.223 | Y | 1.625 | Eu | 1.620 | Fr | 2.567 |
| S | 1.293 | Zr | 1.463 | Gd | 1.623 | Ra | 2.114 |
| Cl | 1.431 | Nb | 1.362 | Tb | 1.613 | Ac | 1.838 |
| Ar | 1.933 | Mo | 1.294 | Dy | 1.613 | Th | 1.655 |
| K | 2.151 | Tc | 1.257 | Ho | 1.604 | Pa | 1.436 |
| Ca | 1.761 | Ru | 1.249 | Er | 1.602 | U | 1.339 |
| Sc | 1.466 | Rh | 1.264 | Tm | 1.602 | Np | 1.291 |
| Ti | 1.308 | Pd | 1.306 | Yb | 1.759 | Pu | 1.271 |
| V | 1.209 | Ag | 1.379 | Lu | 1.605 | Am | 1.261 |
| Cr | 1.162 | Cd | 1.509 | Hf | 1.454 | Cm | 1.279 |

Looking at Fig. 1, one can see a clear correlation between these properties, which means that one of them, or better still, some combination of the two, can be approximately used as a single parameter approximately characterizing chemistry of the element. Therefore, we computed the regression line of the elements in the $\chi$ (Pauling electronegativity) and $R$ (atomic radius) space (Fig. 1), and projected all the elements on it. The chemical scale equal to zero is assigned to the projection of the first element on the regression line, and the distance of the projections of other elements from this "zero" represents their chemical scale. The Mendeleev number - i.e. the integer version of the chemical scale, - is obtained as the sequence of the projected elements along the regression line. The redefined MN is shown in the Table 2.

Table 2. The redefined MN using our method for the first 96 elements of the Periodic Table.

| 1-20 | Fr | Cs | Rb | K | Ra | Ba | Sr | Ac | Ca | Na | Rn | Yb | La | Pm | Tb | Sm | Gd | Eu | Y | Dy |
|---|---|---|---|---|---|---|---|---|---|---|---|---|---|---|---|---|---|---|---|---|
| 21-40 | Th | Ho | Er | Tm | Lu | Li | Ce | Mg | Pr | Hf | Xe | Zr | Nd | Sc | Tl | Pa | Pu | U | Cm | Am |
| 41-60 | Np | Cd | Pb | Ta | In | Po | At | Nb | Ti | Al | Bi | Sn | Zn | Hg | Te | Sb | Ga | V | Mn | Ag |
| 61-80 | Cr | Be | Kr | Ge | Re | Si | Tc | Cu | I | Fe | As | Ni | Co | Mo | Ar | Pd | Ir | Os | Pt | Ru |
| 81-96 | P | Rh | W | Se | Au | B | S | Br | Cl | H | Ne | He | C | N | O | F | | | | |

## Input parameters

Pareto optimization of the hardness and stability was performed over all possible structures (with up to 12 atoms in the primitive cell) in the space of all possible binary compounds formed by 74 elements (i.e. excluding noble gases, rare earth elements and elements heavier than Pu). The input parameters of MendS were: Population size=30, percentage of chemical heredity=30%, percentage of reactive heredity=30%, percentage of mutation=20% and percentage of random systems=20%.

In each binary system, we ran evolutionary optimization with parameters: initial population size=100, subsequent population size=50, number of generations=6, percentage of heredity=40%, percentage of mutation=30%, percentage of random selection=30%.

The comprehensive multi-objective evolutionary algorithm runs as implemented in USPEX was performed on the selected promising binary systems with the following input parameters: initial population size=120, subsequent population size=60, number of generations=50, percentage of heredity=40%, percentage of transmutation=20%, percentage of softmutation=20%, percentage of random selection=20%.

The underlying ab initio structure relaxations and energy calculations were performed using density functional theory with projector augmented wave method (PAW) as implemented in the VASP code[3,4]. In spin-polarized calculations, the GGA-PE functional[5] was implemented. The cutoff energy of 600 eV, and k-mesh with the resolution of $0.06 \times Å^{-1}$ was used in all the VASP calculations. In 20 coevolutionary generations, 600 binary systems were studied. This calculation was done within 3 weeks using 280 cores, a surprisingly minor cost for a full exploration of the chemical space.

## More details on the discussed systems

The crystal structure of several promising compounds presented in the Table. 1 of the paper, were shown in the Fig. 5a in the paper, and here, in Fig. 2, we show the crystal structure of the rest of the discussed compounds in this Table (Table. 1 in the paper).

We also show the convex hull diagram (Fig. 3) constructed during precise evolutionary search on some these binary systems – systems discussed in the Table. 1 of the paper.

In Table. 2, we show a full list of the studied systems during our MendS calculation. One can see that almost all the hard systems, or systems that reported to have a potential to be hard are found and studied in this single run.

## Technicalities

(1) MendS works for all Unary, Binary, Ternary and Quaternary systems. (2) MendS can search for best systems at arbitrary pressures, because atomic radii and electronegativity can be computed for different pressures. (3) Calculations are highly parallelized and run efficiently on modern clusters and supercomputers.

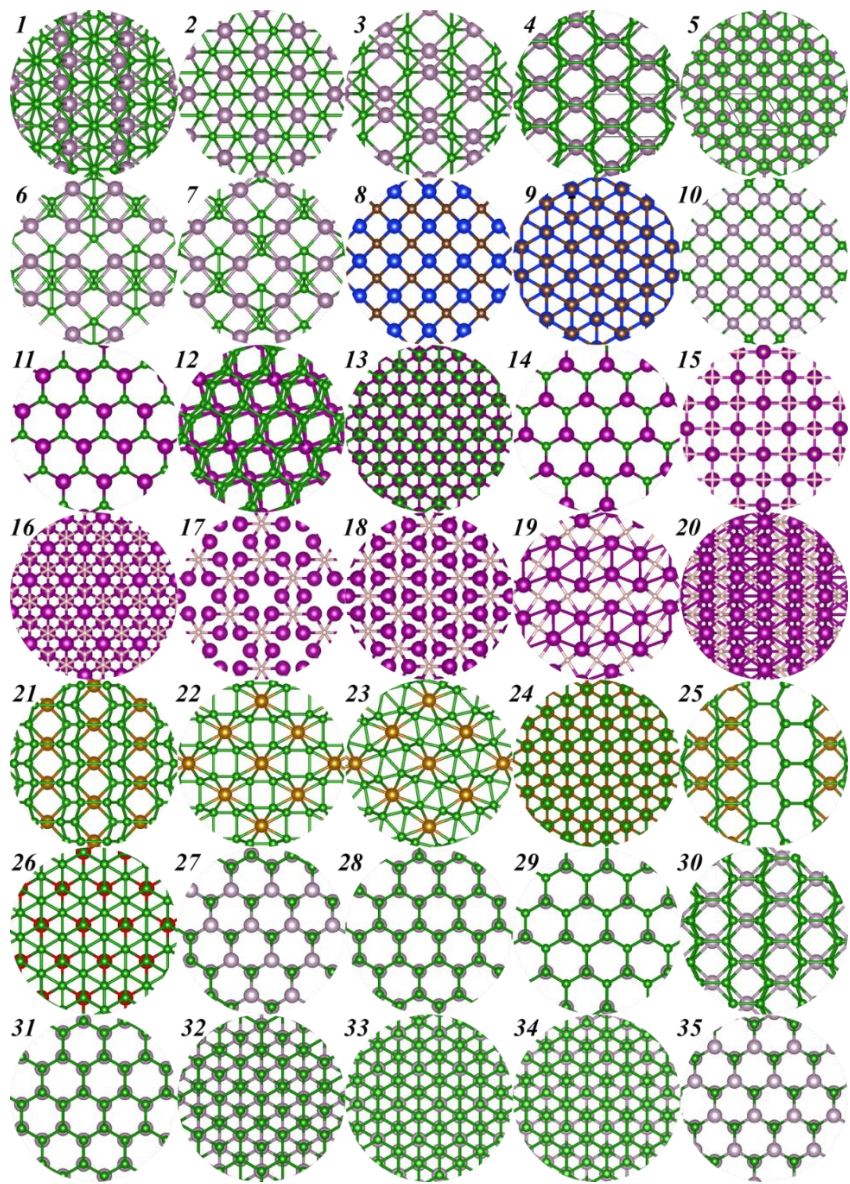

Fig 2. (1) A2/m-MoB$_3$, (2) P6$_3$/mmc-MoB$_3$, (3) R$\bar{3}$m-MoB$_3$, (4) Pmmn-MoB$_4$, (5) R3m-MoB$_8$, (6) Cmcm-Mo$_2$B$_3$, (7) Imm2-Mo$_2$B$_3$, (8) F$\bar{4}$3m-SiC, (9) R3m-SiC, (10) F$\bar{4}$3m-BP, (11) P$\bar{6}$m2-MnB$_3$, (12) P2$_1$/c-MnB$_4$, (13) R$\bar{3}$m-MnB$_4$, (14) P$\bar{6}$m2-MnB$_5$, (15) Fm$\bar{3}$m-MnH, (16) R$\bar{3}$m-MnH, (17) P6$_3$/mcm-Mn$_3$H$_2$, (18) R32-Mn$_3$H$_2$, (19) P2/m-Mn$_4$H$_3$, (20) A2/m-Mn$_6$H$_5$, (21) A2/m-FeB$_4$, (22) Immm-FeB$_4$, (23) Pnnm-FeB$_4$, (24) R$\bar{3}$m-FeB$_4$, (25) Pm-Fe$_2$B$_{11}$, (26) P3m1-VB$_7$, (27) P$\bar{3}$m1-TcB, (28) P$\bar{3}$m1-TcB$_3$, (29) P$\bar{6}$m2-TcB$_3$, (30) P2$_1$/m-TcB$_4$, (31) P6$_3$mmc-TcB$_4$, (32) R$\bar{3}$m-TcB$_4$, (33) R3m-TcB$_7$, (34) R3m-TcB$_8$, (35) P$\bar{6}$m2-Tc$_3$B$_5$.

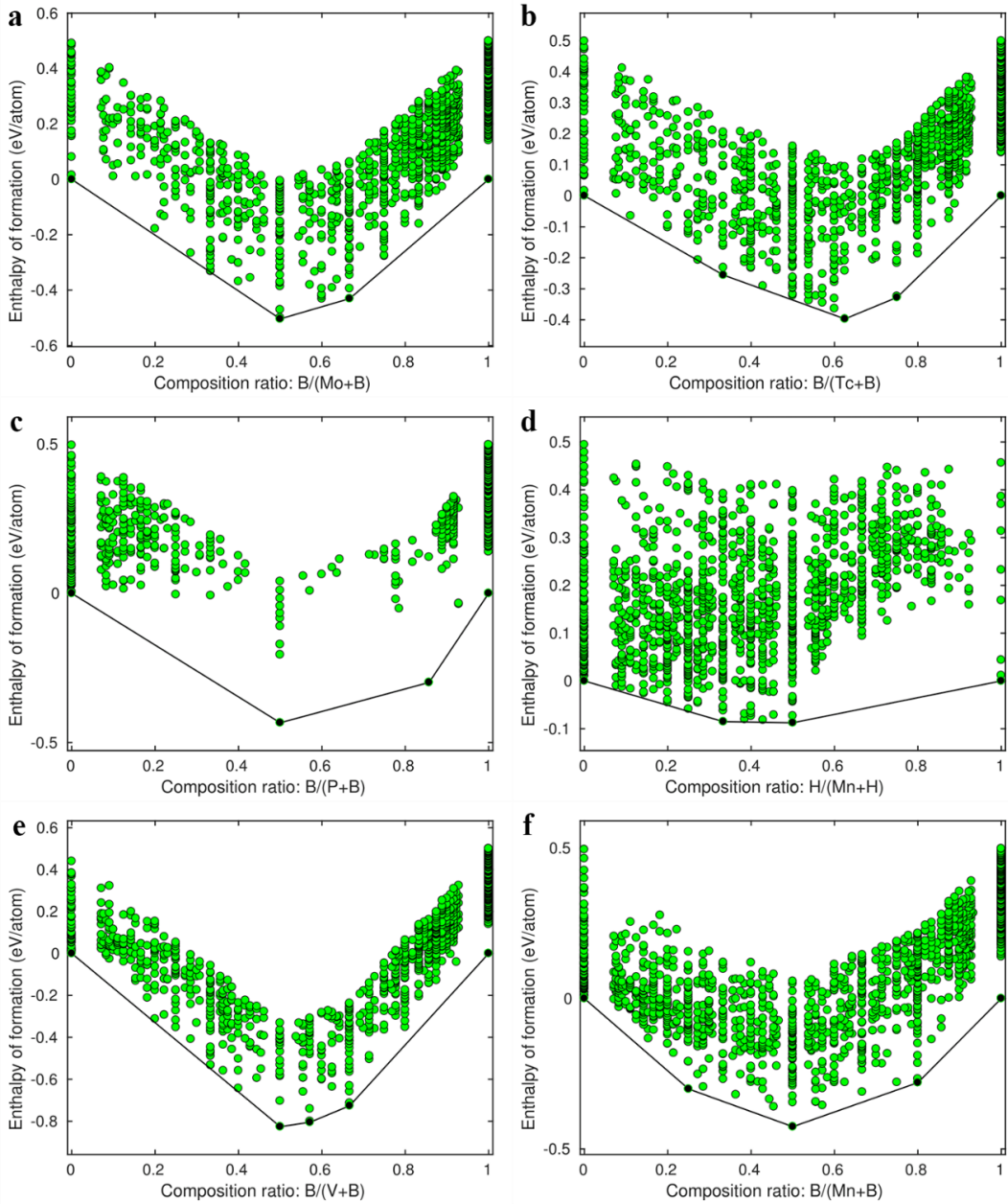

Fig 3. The convex hull plot of discussed systems in the Table. 1 of the paper. (a) Mo-B, (b) Tc-B, (c) B-P, (d) Mn-H, (e) V-B, and (f) Mn-B.

Table 3. Binary systems that were calculated during the MendS run. "How created" specifies the variation operator that was used to create the compounds. In this column R=Randomly, CH=Chemical Heredity, RH=Reactive Heredity, M=Mutation.

| Gen | Compound | How created | 1th Parent | 2nd Parent | Gen | Compound | How created | 1th Parent | 2nd Parent |
|---|---|---|---|---|---|---|---|---|---|
| 1 | Rh-B | R | - | - | 2 | Cr-Co | CH | B | Mo |
| 1 | Na-B | R | - | - | 2 | Rh-W | CH | C | Nb |
| 1 | Ca-Po | R | - | - | 2 | Rh-Rh | CH | Mo | C |
| 1 | Zr-Ge | R | - | - | 2 | Os-P | CH | Mo | Pt |
| 1 | H-N | R | - | - | 2 | Al-Si | CH | Mo | Np-Co |
| 1 | Y-N | R | - | - | 2 | V-Cr | CH | Os | Be |
| 1 | Ga-H | R | - | - | 2 | P-P | CH | Mo | Ru |
| 1 | La-Pt | R | - | - | 2 | Ga-S | CH | Nb-N | As |
| 1 | Sn-B | R | - | - | 2 | Re-C | CH | Nb-N | Rh |
| 1 | Ru-N | R | - | - | 2 | Mo-C | RH | C | Mo-Ir |
| 1 | Rb-Be | R | - | - | 2 | Fe-B | RH | Fe | B |
| 1 | W-C | R | - | - | 2 | As-Co | RH | As | Np-Co |
| 1 | Fe-Os | R | - | - | 2 | Al-Os | RH | Os-Ru | Bi-Al |
| 1 | Ti-Pd | R | - | - | 2 | As-Mo | RH | As | Mo |
| 1 | Th-Be | R | - | - | 2 | Co-Rh | RH | Co | Rh |
| 1 | Rb-Mo | R | - | - | 2 | Al-Be | RH | Be | Al-As |
| 1 | Ra-Zr | R | - | - | 2 | Ru-C | RH | Ru | W-C |
| 1 | Ca-Rh | R | - | - | 2 | W-H | RH | W-C | Ga-H |
| 1 | Np-Co | R | - | - | 2 | Pd-W | M | W | - |
| 1 | Rb-H | R | - | - | 2 | Sb-Rh | M | Rh | - |
| 1 | Bi-Al | R | - | - | 2 | Sn-N | M | Ru-N | - |
| 1 | Os-Ru | R | - | - | 2 | Ta-C | M | C | - |
| 1 | Li-Zn | R | - | - | 2 | In-Rh | M | Rh | - |
| 1 | Al-As | R | - | - | 2 | Mn-Be | M | Be | - |
| 1 | Mo-Ir | R | - | - | 2 | Pu-Ge | R | - | - |
| 1 | Zr-Fe | R | - | - | 2 | Li-Ir | R | - | - |
| 1 | Ac-Zn | R | - | - | 2 | Li-Fe | R | - | - |
| 1 | Nb-N | R | - | - | 2 | Ta-Po | R | - | - |
| 1 | Hf-Si | R | - | - | 2 | Sr-P | R | - | - |
| 1 | Pa-Ir | R | - | - | 2 | P-Cl | R | - | - |
| 3 | As-P | CH | C | F-B | 4 | Be-Fe | CH | C | Mn |
| 3 | Fe-Co | CH | C | C-Cr | 4 | Fe-Fe | CH | Cr | B-H |
| 3 | B-H | CH | Fe | C | 4 | Pd-Au | CH | As | Mo-H |
| 3 | Mo-H | CH | W-H | B | 4 | Cu-P | CH | V-Mo | Ru |
| 3 | Cu-Mo | CH | Ru | Co-Rh | 4 | Pb-Nb | CH | Cr | Fe |
| 3 | Ga-Os | CH | B | C | 4 | Ge-Ge | CH | B | Sn |
| 3 | Co-Ni | CH | V | B | 4 | Cr-Cr | CH | Fe | Mn |
| 3 | Cr-Cu | CH | Mn | Os | 4 | Ti-Ru | CH | Cr-Cu | Cr-Rh |
| 3 | Cu-Os | CH | Ir | Co-Rh | 4 | P-Pt | CH | As-P | Mo-H |
| 3 | Mn-B | RH | Mn | B | 4 | As-B | RH | B | As |
| 3 | V-P | RH | V | Os-P | 4 | B-C | RH | C | B-H |
| 3 | Pd-Os | RH | Pd | Al-Os | 4 | Fe-Cl | RH | Fe | Cl |
| 3 | V-Mo | RH | V | Mo-C | 4 | Np-Mn | RH | Mn | Np |

| Gen | Compound | How created | 1th Parent | 2nd Parent | Gen | Compound | How created | 1th Parent | 2nd Parent |
|---|---|---|---|---|---|---|---|---|---|
| 3 | Cr-Ir | RH | Cr | Ir | 4 | Ca-C | RH | Ca | C |
| 3 | Sn-C | RH | Sn | C | 4 | Mn-P | RH | Mn | As-P |
| 3 | Cr-Rh | RH | Cr | Co-Rh | 4 | Pb-Mn | RH | Mn | Pb |
| 3 | Mn-Pd | RH | Mn | Pd-W | 4 | V-Ir | RH | V | Cr-Ir |
| 3 | As-C | RH | Mo-C | As-Mo | 4 | P-B | RH | B | P |
| 3 | Pt-Rh | M | Co-Rh | - | 4 | Cd-Mn | M | Mn | - |
| 3 | Se-C | M | Mo-C | - | 4 | Mn-C | M | C | - |
| 3 | Ge-W | M | Pd-W | - | 4 | Ge-Cr | M | Cr | - |
| 3 | Rh-C | M | Ru-C | - | 4 | Ge-C | M | C | - |
| 3 | Se-W | M | W-H | - | 4 | Fe-Rh | M | Fe | - |
| 3 | Mo-Ru | M | Mo | - | 4 | Ti-V | M | V | - |
| 3 | Li-Ta | R | - | - | 4 | V-V | R | - | - |
| 3 | Po-P | R | - | - | 4 | Ru-Rh | R | - | - |
| 3 | Sr-Cl | R | - | - | 4 | Bi-As | R | - | - |
| 3 | Zn-P | R | - | - | 4 | Mn-Rh | R | - | - |
| 3 | Al-Sb | R | - | - | 4 | Nb-C | R | - | - |
| 3 | Pb-V | R | - | - | 4 | Ge-Au | R | - | - |
| 5 | Al-Ru | CH | C | B | 6 | Fe-Se | CH | Cr-C | B-C |
| 5 | Be-Be | CH | Mn | B-C | 6 | Pd-Ir | CH | Fe | C |
| 5 | Fe-Ni | CH | Cr | B-C | 6 | Nb-Ga | CH | Cr-B | Cr-C |
| 5 | As-H | CH | Nb-C | B-H | 6 | Fe-Ru | CH | B | Mn-S |
| 5 | Zn-Cu | CH | V | Mn-Rh | 6 | Tc-Pt | CH | C | Co |
| 5 | Hg-Fe | CH | B | B-C | 6 | Al-P | CH | Mn-Pt | Be-B |
| 5 | Mo-P | CH | Cr | Mn-Rh | 6 | Cu-Cu | CH | P | Co |
| 5 | Si-Cu | CH | Be | Fe | 6 | Ru-H | CH | Mn-S | Cr |
| 5 | Co-Co | CH | Cr | B | 6 | Cu-Co | CH | Fe | Ni |
| 5 | Cr-C | RH | B-C | Ge-Cr | 6 | Mo-B | RH | B-H | Mo |
| 5 | P-H | RH | B-H | P-B | 6 | Cu-H | RH | P-H | Cu |
| 5 | Be-B | RH | B | Be | 6 | Fe-Mo | RH | Fe | Mo |
| 5 | Cr-B | RH | Ge-Cr | Mn-B | 6 | Ni-P | RH | Fe-Ni | Fe-P |
| 5 | Fe-Ir | RH | Fe | V-Ir | 6 | Si-B | RH | Si | B-H |
| 5 | Be-Rh | RH | Mn-Rh | Be | 6 | Cr-P | RH | P | Cr |
| 5 | Mn-Cu | RH | Mn | Cu-P | 6 | Fe-S | RH | Fe-Ni | Mn-S |
| 5 | Fe-P | RH | Fe | Cu-P | 6 | Mn-H | RH | P-H | Mn-Pt |
| 5 | Ir-H | RH | V-Ir | B-H | 6 | Mn-As | RH | Mn | As |
| 5 | Mo-Pd | M | Pd | - | 6 | Ge-Fe | M | Fe-Ir | - |
| 5 | Mn-S | M | Mn-C | - | 6 | Tc-Ni | M | Fe-Ni | - |
| 5 | Pd-Rh | M | Rh | - | 6 | Os-S | M | Mn-S | - |
| 5 | Sc-Nb | M | Nb | - | 6 | H-H | M | B-H | - |
| 5 | V-Ru | M | V-Ir | - | 6 | U-Mn | M | Mn | - |
| 5 | Mn-Pt | M | Mn-Rh | - | 6 | Be-Pt | M | Mn-Pt | - |
| 5 | K-Ti | R | - | - | 6 | Ca-Be | R | - | - |
| 5 | Hg-Ag | R | - | - | 6 | Mo-S | R | - | - |
| 5 | Po-Be | R | - | - | 6 | In-Br | R | - | - |
| 5 | Pu-Sn | R | - | - | 6 | Sc-Co | R | - | - |
| 5 | Cs-O | R | - | - | 6 | Cr-N | R | - | - |
| 5 | Ca-Th | R | - | - | 6 | Ga-V | R | - | - |

| Gen | Compound | How created | 1th Parent | 2nd Parent | Gen | Compound | How created | 1th Parent | 2nd Parent |
|---|---|---|---|---|---|---|---|---|---|
| 7 | Ni-B | CH | Cr-N | B-C | 8 | Ni-C | CH | Si-H | B-H |
| 7 | Co-Os | CH | Mo-S | B | 8 | Zn-Zn | CH | Ti | Tc |
| 7 | Os-Pt | CH | Mo-S | B | 8 | Ni-Ni | CH | Co | Fe-Ru |
| 7 | Zn-Fe | CH | Si-B | Cr | 8 | Al-H | CH | Mn-Fe | Cr-N |
| 7 | Ga-Ru | CH | Cr-P | Mn-H | 8 | Ge-Cu | CH | Te | Cr-B |
| 7 | Al-Cr | CH | Ru-H | Mn-H | 8 | Po-Fe | CH | Mn-Fe | C |
| 7 | Ni-Ru | CH | Tc-Pt | B | 8 | Tc-Br | CH | Cr-N | Te |
| 7 | Be-Co | CH | B-H | Mn | 8 | Ti-Ti | CH | Al-Cr | Ti-Cr |
| 7 | Co-Ru | CH | C | Cr-P | 8 | Tc-Ir | CH | Mn-C | Mo |
| 7 | Si-H | RH | Si | Mn-H | 8 | Mn-Tc | RH | Mn | Tc |
| 7 | Tc-P | RH | P | Tc | 8 | Cr-H | RH | Mn-H | Mn-Cr |
| 7 | Cr-Mo | RH | Mo-S | Cr | 8 | Tc-Fe | RH | Tc | Fe |
| 7 | S-B | RH | Mo-S | Si-B | 8 | Os-H | RH | Mn-H | Os-Pt |
| 7 | Tc-B | RH | B-C | Tc | 8 | Cr-Fe | RH | Cr-N | Mn-Fe |
| 7 | Co-B | RH | B-C | Co | 8 | Cr-Os | RH | Os | Mn-Cr |
| 7 | Mn-Cr | RH | Cr-N | Mn-H | 8 | Mn-N | RH | Ti-Mn | Cr-N |
| 7 | Mn-Co | RH | Co | Mn-As | 8 | Ti-Be | RH | Be | Ti-Cr |
| 7 | Mn-Fe | RH | Mn-As | Fe | 8 | Pd-B | RH | B-H | Pd |
| 7 | Fe-H | M | H | - | 8 | Zn-Mo | M | Cr-Mo | - |
| 7 | Tl-Ru | M | Ru-H | - | 8 | B-B | M | Tc-B | - |
| 7 | Ti-Cr | M | Cr | - | 8 | Sb-Ni | M | Ni | - |
| 7 | Te-Mn | M | Mn | - | 8 | V-Tc | M | Tc-B | - |
| 7 | Al-Pt | M | Al-P | - | 8 | Be-Ni | M | Be | - |
| 7 | Ti-Mn | M | Mn-As | - | 8 | Mn-Mo | M | Cr-Mo | - |
| 7 | Ca-Ni | R | - | - | 8 | Ac-Si | R | - | - |
| 7 | Zn-Be | R | - | - | 8 | Hf-Bi | R | - | - |
| 7 | Br-S | R | - | - | 8 | Pa-Si | R | - | - |
| 7 | In-Co | R | - | - | 8 | Pu-Tc | R | - | - |
| 7 | Rb-Sc | R | - | - | 8 | U-Se | R | - | - |
| 7 | Tc-Tc | R | - | - | 8 | Pd-O | R | - | - |
| 9 | P-C | CH | Os-H | B-H | 10 | Pa-P | CH | Rh | B-H |
| 9 | Co-Ir | CH | Mn-C | C | 10 | Si-Fe | CH | Mn | Cr-Ni |
| 9 | Ni-O | CH | Cr-Fe | Mn-N | 10 | Co-P | CH | V-B | B-H |
| 9 | Tc-Cu | CH | Mo | Cr | 10 | In-Ge | CH | Cr-B | Cr-H |
| 9 | Ga-Be | CH | Cr-B | Mn | 10 | V-Ni | CH | Tc-Ru | B |
| 9 | Cu-Ru | CH | Cr-H | Os | 10 | Ge-Ni | CH | Cr | Cr-N |
| 9 | Hg-Ni | CH | C | Cr-B | 10 | Hf-Be | CH | Mn | Cr-H |
| 9 | Cd-Be | CH | C | Mn-H | 10 | Nb-Co | CH | Cr | Si |
| 9 | Cu-Rh | CH | Cr-N | Os | 10 | Np-Cr | CH | Po | Fe |
| 9 | V-B | RH | V | B | 10 | C-O | RH | Ni-O | Tc-C |
| 9 | Ni-N | RH | Cr-N | Ni | 10 | Co-N | RH | Cr-N | Co |
| 9 | Mn-Ni | RH | Ni-C | Mn-H | 10 | Si-Tc | RH | Si | Tc-C |
| 9 | Cr-Ni | RH | Ni-C | Cr-Fe | 10 | Tc-H | RH | B-H | Tc |
| 9 | C-H | RH | B-H | C | 10 | Ga-C | RH | Ga | C |
| 9 | Tc-C | RH | C | Tc | 10 | Nb-V | RH | Nb-Ga | V-B |
| 9 | Ir-N | RH | Ir | Cr-N | 10 | Ga-Ni | RH | Ni | Ga-Be |
| 9 | Tc-Po | RH | Tc | Po | 10 | Be-N | RH | Be | Cr-N |

| Gen | Compound | How created | 1th Parent | 2nd Parent | Gen | Compound | How created | 1th Parent | 2nd Parent |
|---|---|---|---|---|---|---|---|---|---|
| 9 | C-N | RH | Mn-N | C | 10 | Po-H | RH | B-H | Po |
| 9 | Zn-Mn | M | Mn | - | 10 | Re-Fe | M | Fe | - |
| 9 | Co-H | M | B-H | - | 10 | Zr-Be | M | Be | - |
| 9 | Cu-C | M | B-C | - | 10 | V-Re | M | V | - |
| 9 | V-Fe | M | Fe | - | 10 | Ga-Tc | M | Ga-Be | - |
| 9 | Tc-Ru | M | Tc-Ir | - | 10 | Cr-W | M | Cr-N | - |
| 9 | Si-C | M | Ni-C | - | 10 | Re-Cu | M | Cu | - |
| 9 | Y-Np | R | - | - | 10 | Sr-U | R | - | - |
| 9 | Li-Sn | R | - | - | 10 | Y-Zr | R | - | - |
| 9 | Sr-I | R | - | - | 10 | Al-Mn | R | - | - |
| 9 | Te-Co | R | - | - | 10 | Sb-Au | R | - | - |
| 9 | Rb-Ba | R | - | - | 10 | Ag-C | R | - | - |
| 9 | Hg-Be | R | - | - | 10 | Tl-Fe | R | - | - |
| 11 | V-Rh | CH | Mn | Al-Mn | 12 | P-N | CH | Br-O | B-H |
| 11 | Mo-Ni | CH | V-Re | Mn-H | 12 | Hg-Os | CH | B-H | B-C |
| 11 | Po-Cu | CH | V-Ni | C | 12 | Be-F | CH | Mn-O | O |
| 11 | Cd-Pt | CH | Ga-C | Np-Cr | 12 | Mo-Mo | CH | P | Si |
| 11 | Al-Co | CH | B | Ga-C | 12 | Mg-Fe | CH | Al-C | Zn-Cr |
| 11 | Mg-In | CH | B-C | Ni | 12 | Re-Ru | CH | Zn-Cr | Si |
| 11 | Zn-Cr | CH | V | Tc-H | 12 | Ta-Be | CH | Cr | Be |
| 11 | C-C | CH | B-H | O | 12 | Co-C | CH | B-H | Cr-O |
| 11 | Co-Mo | CH | Cr-W | Ni | 12 | Be-C | CH | B-H | Be-O |
| 11 | Al-C | RH | B-C | Al-Mn | 12 | Cr-Be | RH | Be | Cr-N |
| 11 | Cr-O | RH | O | Cr | 12 | Zn-V | RH | Zn-Cr | V |
| 11 | V-Si | RH | V | Si | 12 | B-O | RH | Mn-O | Cr-B |
| 11 | Mn-O | RH | Mn | O | 12 | Ga-Mo | RH | Mo | Ga |
| 11 | V-Mn | RH | Al-Mn | V | 12 | N-O | RH | O | Cr-N |
| 11 | V-Be | RH | Zr-Be | V | 12 | Be-Br | RH | Br | V-Be |
| 11 | Fe-N | RH | Fe | Cr-N | 12 | Nb-Cu | RH | Nb-Ga | Cu |
| 11 | Cr-Tc | RH | Tc | Cr-N | 12 | Ga-B | RH | Ga | B-C |
| 11 | Be-O | RH | Be | O | 12 | Zn-O | RH | O | Zn-Cr |
| 11 | Br-O | M | O | - | 12 | Pd-N | M | Fe-N | - |
| 11 | Tl-Mn | M | Mn | - | 12 | Ti-B | M | B-H | - |
| 11 | Pu-N | M | Cr-N | - | 12 | Si-O | M | Cr-O | - |
| 11 | Ge-B | M | B | - | 12 | Nb-Tc | M | Cr-Tc | - |
| 11 | Tc-Co | M | Tc-H | - | 12 | Pa-Be | M | Be | - |
| 11 | Tc-W | M | Tc | - | 12 | Si-S | M | Fe-S | - |
| 11 | Pu-Cu | R | - | - | 12 | Np-Os | R | - | - |
| 11 | Ba-Co | R | - | - | 12 | Te-Ru | R | - | - |
| 11 | Hf-Pa | R | - | - | 12 | Rb-Li | R | - | - |
| 11 | Cu-W | R | - | - | 12 | Y-Nb | R | - | - |
| 11 | Nb-P | R | - | - | 12 | Sc-I | R | - | - |
| 11 | Fr-Cu | R | - | - | 12 | Tl-Pd | R | - | - |
| 13 | N-N | CH | C | O | 14 | V-C | CH | Cr-B | Cr-N |
| 13 | P-Rh | CH | Ti | O | 14 | Ga-Rh | CH | Se | Mn-H |
| 13 | Pu-S | CH | Cr-N | Nb-Tc | 14 | Cu-Ni | CH | Ga | B |
| 13 | Sn-Ru | CH | Cr-B | Ga-Mo | 14 | Zr-Zn | CH | Nb-Mn | Cr |

| Gen | Compound | How created | 1th Parent | 2nd Parent | Gen | Compound | How created | 1th Parent | 2nd Parent |
|---|---|---|---|---|---|---|---|---|---|
| 13 | Re-Tc | CH | Ti | Mo | 14 | Pd-Ru | CH | P-Rh | C |
| 13 | Rh-O | CH | Zn-O | F | 14 | Ni-Ir | CH | B-N | P-Rh |
| 13 | Ru-Ru | CH | P | F | 14 | S-N | CH | O | Fe |
| 13 | In-Ru | CH | Si-S | C | 14 | Ti-Mo | CH | Cr | Fe |
| 13 | Au-C | CH | Fe | B-C | 14 | Zn-Tc | CH | Mo | V |
| 13 | P-O | RH | B-O | P | 14 | V-N | RH | Al-V | Cr-N |
| 13 | B-N | RH | P-N | Cr-B | 14 | Nb-Fe | RH | Nb-Mn | Fe-Se |
| 13 | Cr-Se | RH | Cr-Be | Fe-Se | 14 | As-O | RH | As | P-O |
| 13 | Cr-Ru | RH | Cr | Re-Ru | 14 | C-F | RH | C | F |
| 13 | Nb-Mo | RH | Nb-Ga | Ga-Mo | 14 | Nb-H | RH | Nb-Mn | Mn-H |
| 13 | Be-Mo | RH | Mo | Be-C | 14 | Nb-O | RH | Nb-Mn | O |
| 13 | Fe-Br | RH | Fe | Br | 14 | V-O | RH | Rh-O | V |
| 13 | Re-B | RH | Re | B | 14 | Ru-B | RH | Cr-B | Cr-Ru |
| 13 | Si-P | RH | Si | P | 14 | N-F | RH | B-N | F |
| 13 | Al-V | M | V | - | 14 | S-O | M | O | - |
| 13 | Pa-O | M | Si-O | - | 14 | Ta-Mn | M | Mn | - |
| 13 | Nb-Mn | M | Mn | - | 14 | Zn-N | M | B-N | - |
| 13 | Mn-I | M | Mn | - | 14 | Ir-O | M | P-O | - |
| 13 | Re-H | M | Mn-H | - | 14 | Si-N | M | Tc-N | - |
| 13 | Tc-N | M | P-N | - | 14 | Cu-B | M | Cr-B | - |
| 13 | Ac-C | R | - | - | 14 | Ac-Tl | R | - | - |
| 13 | Ga-As | R | - | - | 14 | Sb-I | R | - | - |
| 13 | Ge-As | R | - | - | 14 | Sb-Be | R | - | - |
| 13 | Ag-Ni | R | - | - | 14 | Pa-Tc | R | - | - |
| 13 | Hf-F | R | - | - | 14 | Ge-Si | R | - | - |
| 13 | Ca-Ac | R | - | - | 14 | U-Au | R | - | - |
| 15 | Ni-Br | CH | Si-N | N-F | 16 | Fe-C | CH | Mo-N | B |
| 15 | Ni-F | CH | Si-N | B-N | 16 | Y-Co | CH | Mn-H | Cr |
| 15 | Mo-N | CH | As-O | Ni-P | 16 | Ni-Os | CH | B-N | Si |
| 15 | I-Fe | CH | O | V-N | 16 | U-Co | CH | Cr-N | Cr |
| 15 | Cu-S | CH | Ta-Mn | F | 16 | Re-Au | CH | Mo | V-Ge |
| 15 | Si-Rh | CH | Pd | V-O | 16 | Tc-Mo | CH | B-C | Mo-N |
| 15 | Ag-W | CH | Nb-Fe | Cr-N | 16 | Mn-Au | CH | C | B-N |
| 15 | Pt-Au | CH | C | Ni | 16 | Ni-S | CH | B-N | Fe |
| 15 | Fe-O | CH | Cr-N | F | 16 | Zn-Si | CH | V | Ru |
| 15 | V-F | RH | N-F | V-C | 16 | Cr-Pt | RH | Pt | Cr-B |
| 15 | Ru-O | RH | V-O | Ru-B | 16 | As-Ir | RH | Ir | As |
| 15 | Zr-V | RH | Zr | V-N | 16 | Cr-Si | RH | Cr-N | Si |
| 15 | H-F | RH | Mn-H | N-F | 16 | Si-F | RH | Si | F |
| 15 | Tc-F | RH | F | Tc | 16 | B-F | RH | H-F | B-C |
| 15 | Cr-F | RH | N-F | Cr-N | 16 | Mo-O | RH | O | Mo-N |
| 15 | V-Ge | RH | V | Ge | 16 | Mn-Ru | RH | Mn-H | Ru |
| 15 | As-Ni | RH | As-O | Ni | 16 | As-F | RH | F | Cr-As |
| 15 | Cr-As | RH | As | Cr | 16 | Zr-O | RH | Zr-V | O |
| 15 | Bi-V | M | V-C | - | 16 | V-Ag | M | V | - |
| 15 | Si-Ir | M | Si-N | - | 16 | Nb-Si | M | Si | - |
| 15 | Cd-Ga | M | Nb-Ga | - | 16 | Pu-Ni | M | Ni | - |

| Gen | Compound | How created | 1th Parent | 2nd Parent | Gen | Compound | How created | 1th Parent | 2nd Parent |
|---|---|---|---|---|---|---|---|---|---|
| 15 | Mg-C | M | V-C | - | 16 | O-F | M | F | - |
| 15 | Hg-F | M | N-F | - | 16 | Tc-As | M | Cr-As | - |
| 15 | Pu-H | M | Mn-H | - | 16 | Ir-B | M | B-N | - |
| 15 | Sr-Co | R | - | - | 16 | K-Sb | R | - | - |
| 15 | Ag-Mn | R | - | - | 16 | Li-Cu | R | - | - |
| 15 | Ga-At | R | - | - | 16 | In-S | R | - | - |
| 15 | Ti-Cu | R | - | - | 16 | Ac-F | R | - | - |
| 15 | Ag-Pd | R | - | - | 16 | Pu-F | R | - | - |
| 15 | U-Si | R | - | - | 16 | Hg-Te | R | - | - |
| 17 | Ga-O | CH | Fe-C | Cr-N | 18 | Ta-Ru | CH | Si-Mo | Cr-N |
| 17 | Zn-Co | CH | B-H | Cr-B | 18 | Bi-Fe | CH | Cr-N | B-C |
| 17 | Ni-Rh | CH | Si | B-C | 18 | Ag-Fe | CH | C | B-C |
| 17 | Cd-Tc | CH | Cr-Si | Mo | 18 | Re-P | CH | Cr-Pd | Mo |
| 17 | Fe-Pt | CH | O | Cr-N | 18 | Mg-As | CH | B-C | Nb-B |
| 17 | Rh-N | CH | C | Mo-O | 18 | Al-Fe | CH | Mn | Cr |
| 17 | Al-Mo | CH | Tc | Tc-As | 18 | Co-S | CH | B-C | Cr-Au |
| 17 | I-S | CH | Mo-O | B-N | 18 | Zn-Ge | CH | B-N | C |
| 17 | Co-O | CH | Fe-C | F | 18 | Zn-B | CH | Mn-H | Cr |
| 17 | Mn-Si | RH | Mn | Cr-Si | 18 | Tc-Au | RH | Tc | Au |
| 17 | V-H | RH | V | Mn-H | 18 | Al-Rh | RH | Rh | Al-Mo |
| 17 | Mn-F | RH | Mn | O-F | 18 | H-O | RH | O-F | B-H |
| 17 | Ir-C | RH | Fe-C | Ir-B | 18 | Si-Pt | RH | Mn-Si | Pt |
| 17 | Tc-O | RH | O-F | Tc-As | 18 | Nb-Cr | RH | Cr | Nb |
| 17 | Si-Mo | RH | Tc-Mo | Si | 18 | Mo-F | RH | Mo | O-F |
| 17 | As-N | RH | Cr-N | As | 18 | Pd-C | RH | Cr-Pd | B-C |
| 17 | Pt-F | RH | Cr-Pt | F | 18 | Mo-Rh | RH | Mo | Rh |
| 17 | Nb-B | RH | Nb-Si | B-N | 18 | Ga-F | RH | F | Ga-O |
| 17 | Pa-Mn | M | Mn | - | 18 | Hg-Cr | M | Cr | - |
| 17 | Hg-H | M | B-H | - | 18 | Pu-Pt | M | Pt | - |
| 17 | Ag-Tc | M | Tc | - | 18 | Cd-Pd | M | Pd | - |
| 17 | In-C | M | C | - | 18 | Ti-N | M | Cr-N | - |
| 17 | Cr-Au | M | Cr-N | - | 18 | Ag-N | M | B-N | - |
| 17 | Cr-Pd | M | Cr-B | - | 18 | Zr-Mn | M | Mn-H | - |
| 17 | Mg-Be | R | - | - | 18 | Ta-Mo | R | - | - |
| 17 | Ba-Hg | R | - | - | 18 | Hf-Ni | R | - | - |
| 17 | Hg-Ir | R | - | - | 18 | U-Cu | R | - | - |
| 17 | Bi-At | R | - | - | 18 | Ga-Po | R | - | - |
| 17 | Pa-Sb | R | - | - | 18 | Li-Tl | R | - | - |
| 17 | Ru-Br | R | - | - | 18 | Os-Se | R | - | - |
| 19 | Tc-Rh | CH | Pd | Cr-N | 20 | Re-I | CH | Br | Cr-N |
| 19 | S-Cl | CH | Ni | O | 20 | Be-Cu | CH | Mn | Os-B |
| 19 | Th-Rh | CH | C-C | Mo-Rh | 20 | I-Os | CH | Re | Tc-Rh |
| 19 | Ir-Rh | CH | Se | P | 20 | Bi-Ni | CH | Tc-Rh | Os-B |
| 19 | Au-S | CH | Se | C | 20 | Be-S | CH | Mn-H | B-H |
| 19 | Br-C | CH | O-F | B-N | 20 | Ta-Co | CH | Zr-Cr | Zr-Ni |
| 19 | Ti-C | CH | B-H | B-N | 20 | Sn-Al | CH | Zr-Cr | Re-Re |
| 19 | Zr-Nb | CH | Cr | Mn | 20 | Ag-Mo | CH | As | Mn-Re |

| Gen | Compound | How created | 1th Parent | 2nd Parent | Gen | Compound | How created | 1th Parent | 2nd Parent |
|---|---|---|---|---|---|---|---|---|---|
| 19 | Rh-H | CH | B-N | Mo | 20 | Cl-H | CH | B | C-C |
| 19 | Mn-Re | RH | Zr-Mn | Re-P | 20 | Zr-C | RH | C | Zr-Ni |
| 19 | Zr-B | RH | Zr-Mn | B | 20 | Os-Rh | RH | Rh | Os-B |
| 19 | Ta-O | RH | O | Ta-Mo | 20 | Rh-Cl | RH | S-Cl | Rh-H |
| 19 | Zr-Ni | RH | Ni | Zr-Mn | 20 | Mn-Os | RH | Os-B | Mn |
| 19 | Zr-Cr | RH | Zr-Mn | Cr-N | 20 | S-F | RH | S | F |
| 19 | As-Rh | RH | As | Rh | 20 | Re-Os | RH | Re | Os-B |
| 19 | Pt-N | RH | Cr-N | Pt | 20 | Si-Os | RH | Os-B | Si |
| 19 | Re-F | RH | Re-P | O-F | 20 | Zr-Rh | RH | Rh | Zr-B |
| 19 | Mn-Se | RH | Se | Mn | 20 | Re-Ir | RH | Re | Ir-Rh |
| 19 | Os-B | M | B | - | 20 | Re-Rh | M | Rh-H | - |
| 19 | Re-Si | M | Re | - | 20 | Pt-B | M | B-N | - |
| 19 | Se-F | M | F | - | 20 | Ti-As | M | Ti-C | - |
| 19 | Bi-Cr | M | Cr-N | - | 20 | Ta-Cr | M | Cr | - |
| 19 | Co-F | M | F | - | 20 | Ti-Si | M | Si | - |
| 19 | Cd-Au | M | Tc-Au | - | 20 | W-B | M | B-C | - |
| 19 | Pu-Ru | R | - | - | 20 | K-Pd | R | - | - |
| 19 | I-N | R | - | - | 20 | At-C | R | - | - |
| 19 | Ac-Be | R | - | - | 20 | At-O | R | - | - |
| 19 | Li-C | R | - | - | 20 | Li-Ti | R | - | - |
| 19 | U-Pt | R | - | - | 20 | Rb-Na | R | - | - |
| 19 | Zn-Os | R | - | - | 20 | Pb-C | R | - | - |